\documentclass[conference]{IEEEtran}
\IEEEoverridecommandlockouts
\usepackage{cite}
\usepackage{amsmath,amssymb,amsfonts}

\usepackage{algorithmic}
\usepackage{graphicx}
\usepackage{textcomp}
\usepackage{xcolor}
\usepackage{graphicx}
\usepackage{subcaption}
\usepackage{orcidlink}

\def\BibTeX{{\rm B\kern-.05em{\sc i\kern-.025em b}\kern-.08em
    T\kern-.1667em\lower.7ex\hbox{E}\kern-.125emX}}
\begin{document}

\title{The Potential of Quantum Techniques for Stock Price Prediction
}


\author{\IEEEauthorblockN{1\textsuperscript{st} Naman Srivastava\orcidlink{0000-0001-7521-6525}}
\IEEEauthorblockA{\textit{IIIT Dharwad}\\
Dharwad, India \\
srinaman2@gmail.com}
\and
\IEEEauthorblockN{2\textsuperscript{nd} Gaurang Belekar \orcidlink{0009-0002-1755-7764}}
\IEEEauthorblockA{\textit{IIIT Dharwad}\\
Dharwad, India \\
belekargaurang@gmail.com}
\and
\IEEEauthorblockN{3\textsuperscript{rd} Neel Shahakar \orcidlink{0000-0003-4059-4256}}
\IEEEauthorblockA{\textit{IIIT Dharwad}\\
Dharwad, India \\
neelshahakar@gmail.com}
\and
\IEEEauthorblockN{4\textsuperscript{th} Aswath Babu H.\orcidlink{0009-0007-8251-3828} }
\IEEEauthorblockA{\textit{IIIT Dharwad}\\
Dharwad, India \\
aswath@iiitdwd.ac.in}

}

\maketitle

\begin{abstract}

We explored the potential applications of various Quantum Algorithms for stock price prediction by conducting a series of experimental simulations using both Classical as well as Quantum Hardware. Firstly, we extracted various stock price indicators, such as Moving Averages (MA), Average True Range (ATR), and Aroon, to gain insights into market trends and stock price movements. Next, we employed Quantum Annealing (QA) for feature selection and Principal Component Analysis (PCA) for dimensionality reduction. Further, we transformed the stock price prediction task essentially into a classification problem. We trained the Quantum Support Vector Machine (QSVM) to predict price movements (whether up or down) and contrasted its performance with classical models and analysed their accuracy on dataset formulated using Quantum Annealing and PCA individually. We focused on stock price prediction and binary classification of stock prices for four different companies, namely Apple, Visa, Johnson and Jonson and Honeywell. We primarily used the real-time stock data of the raw stock prices of these companies. We compared various Quantum Computing techniques with their classical counterparts in terms of accuracy and F-score of the prediction model. Through these experimental simulations, we shed light on the potential advantages and limitations of Quantum Algorithms in stock price prediction and contribute to the growing body of knowledge at the intersection of Quantum Computing and Finance.

\end{abstract}

\begin{IEEEkeywords}
Computing Systems (Quantum); QSVM, Quantum Annealing, Feature Selection, Machine Learning
\end{IEEEkeywords}

\section{Introduction}

The financial markets have been a focal point of intensive research for the past few decades, with a constant pursuit of accurate stock price predictions among investors, traders, and analysts. Quantum Computing realized by harnessing quantum properties has emerged as a potentially revolutionary tool in this domain, wherein complex financial challenges can be addressed with greater efficiency and precision in comparison to classical computing methods. This emerging technology has grabbed significant attention, with the research community and industry experts exploring its applications and potential impact on stock price prediction.

Conventional stock price prediction approaches purely rely on statistical models, machine learning algorithms, and time-series analysis. However, these methods encounter limitations when handling vast historical financial data, intricate market dynamics, and rapidly changing conditions. Quantum Computing offers a distinct computational paradigm, capitalizing on Quantum Mechanics principles like Superposition and Entanglement to explore an exponentially larger solution space compared to Classical Computing. Quantum Algorithms have shown the potential for exponentially faster problem-solving, making them intriguing candidates for financial analysis, including stock price prediction\cite{{emmanoulopoulos2022quantum},{liu2022quantum},{naik2023portfolio}}.

Feature selection is a crucial aspect of building accurate Machine Learning Models. Quantum Algorithms like Quantum Annealing have demonstrated their efficiency in feature selection and dimensional reduction tasks. By identifying relevant features more effectively from extensive financial data sets, Quantum Computing can enhance input data quality, resulting in improved stock price forecasting accuracy.

Additionally, Quantum Computing can address optimization challenges in financial modeling. Stock price prediction often requires finding optimal weights for influencing variables, and Quantum Optimization Algorithms like the Quantum Approximate Optimization Algorithm (QAOA) show promise in handling these tasks more efficiently\cite{rebentrost2018quantum,huggins2019towards,cerezo2021variational}.

 Quantum Neural Networks (QNN) amalgamate the principles of Quantum Mechanics with the architecture of Neural Networks, harnessing the potential for exponential computational speedup and enhanced problem-solving capabilities\cite{schuld2014quest,jia2019quantum,situ2020quantum}.
The applications of QNNs span diverse domains, showcasing their potential to revolutionize industries. Financial modeling benefits from QNNs' ability to process intricate data relationships, enhancing portfolio optimization, risk assessment, and option pricing.
Earlier it has been researched on a comparative analysis of forecasting stock prices using Long Short-Term Memory (LSTM) and Quantum Long Short-Term Memory (QLSTM) models\cite{chen2022quantum,abbas2021power}. LSTM, a prevalent Recurrent Neural Network (RNN) architecture, excels at capturing sequential data's temporal patterns, making it apt for predicting time series like stock prices. In contrast, QLSTM leverages Quantum Computing's distinctive traits to potentially enhance predictive accuracy even further \cite{QLSTM_soft, QNN,takaki2021learning,paquet2022quantumleap}.

QLSTM represents an innovative advancement in the realm of quantum machine learning, blending the power of quantum computing with the capabilities of LSTM networks. However, Quantum Computing faces challenges, including noise and errors with current quantum hardware, which may limit result accuracy. Identifying specific financial analysis tasks where Quantum Computing excels remains an ongoing research area. This study delves into the potential applications of Quantum Computing in stock price prediction and explores its benefits and limitations to contribute to the growing knowledge in this emerging field.

We explored the use of Quantum Computing in stock price prediction, examining various Quantum Algorithms and their potential benefits in financial analysis. We conducted experiments using real-world financial data and compared the performance of quantum approaches with classical methods to evaluate the feasibility and implications of integrating Quantum Computing into the domain of stock market forecasting. Through this investigation, we aim to contribute to the growing body of knowledge on the practical applications of Quantum Computing in finance and its potential impact on stock price prediction methodologies.     

\section{Background}

\subsection{Stock Market Indicators}

The stock market is a complicated and dynamic domain influenced by a multitude of factors. To navigate this intricate landscape and make informed investment decisions, market participants heavily rely on a diverse range of indicators. These indicators are essential tools that aid in analyzing market trends, identifying potential opportunities, and assessing risk levels. Various types of indicators exist, which include trend-following indicators like Moving Averages, that provide insights into price trends over time. Oscillators, such as the Relative Strength Index (RSI), offer signals of overbought or oversold conditions. Volatility indicators, such as the Bollinger Bands, assist in measuring market volatility and further recognize potential price movements.
Additionally, volume-based indicators, such as On-Balance-Volume (OBV), give valuable information on the strength of price trends based on trading volume. In this study, we explored and evaluated the performance of these different indicators in the context of stock market analysis and their potential contributions to stock price prediction and decision-making processes. Understanding the significance and implications of these indicators is crucial for developing effective investment strategies and improving financial decision-making.

These indicators are just a few examples of the many tools available to stock market participants. Traders and investors often use overlapping indicators to gain a comprehensive understanding of the dynamics of the market to plan their trading strategies. However, it is essential to note that no indicator is foolproof, and using multiple indicators can help mitigate risks and enhance decision-making in the highly varying and uncertain stock market environment.

\subsection{Quantum Annealing}
Quantum Annealing (QA) based on Quantum Computing techniques, has emerged as a promising approach for feature selection in various domains. As an optimization technique, QA aims to find the optimal feature subset that maximizes or minimizes a given objective function. In the context of feature selection, QA explores the energy landscape of the feature space, seeking the most relevant features that contribute significantly to the predictive power of a model. Unlike classical feature selection methods, QA can efficiently explore a vast number of possible feature combinations simultaneously, potentially leading to more accurate and efficient feature selection\cite{morita2008mathematical,finnila1994quantum,mucke2023feature}.

\subsection{Quantum Support Vector Classifier}
Quantum Support Vector Machine (QSVM) is a revolutionary extension of the classical Support Vector Machine (SVM) algorithm, that includes the power of quantum computing, in turn, the additional ability to address complex classification problems \cite{rebentrost2014quantum}. With the growing interest in quantum machine learning, QSVM has garnered significant attention for its potential to outperform classical SVM in certain scenarios. Having leveraged using Quantum Computing methods, QSVM facilitates the process of figuring out optimal hyper-planes in high-dimensional feature spaces, leading to more accurate and efficient binary classification tasks\cite{kariya2021investigation}. 


\section{Quantum Enhanced Pipeline}

\begin{figure*}[h]
    \centering
    \includegraphics[scale=0.3]{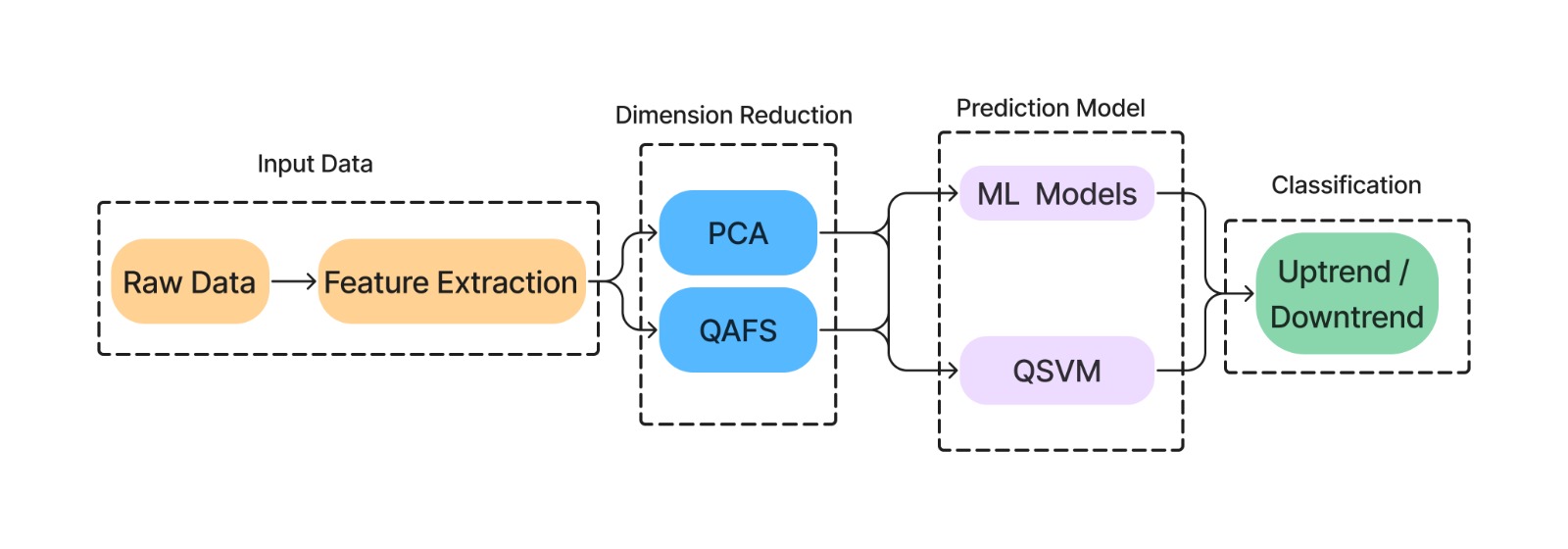}
    \caption{Workflow depicting Quantum Assisted Pipeline for Stock Price Prediction}
    \label{fig:WORKFLOW}
\end{figure*}

\subsection{Raw Data}
We have collected real-time raw data for the stock prices of four companies namely Honeywell (HON), Johnson and Johnson (JNJ), Apple(AAPL) and Visa (VISA) using the yfinance library. Here raw data includes the closing price, Highest and Lowest price for the stock for each given day, from 25 December 2020 to 25 December 2022 extracted through yfinance API. As these companies belong to different domains, which allows us to explore the model's performance on different data patterns. 

\subsection{Feature Extraction}
In the stock market, various indicators are used to analyze and interpret market trends, price movements, and overall market sentiment. These indicators provide valuable insights to traders and investors for making informed decisions. For the available historical data of the stocks, we computed the following indicators, which were employed for undertaking Feature Extraction:\\
1. Moving Averages (MA): Describe the average price of a security over a specified period, wherein price fluctuations are smoothed out and trends are highlighted. Among two kinds of MA, one is Simple Moving Average (SMA) and the other is Exponential Moving Average (EMA).\\
2. Relative Strength Index (RSI): Measures the speed with which change of price movements occurs, and that indicates overbought or oversold conditions. RSI values range from 0 to 100, wherein the readings above 70 are treated as overbought and anything below 30 is oversold.\\
3. Moving Average Convergence Divergence (MACD): It is a trend-following momentum indicator that compares two moving averages of a security's price. It provides signals when the two moving averages converge or diverge.\\
4. Stochastic Oscillator: This momentum indicator compares a closing of security's closing price to its price range over a specific period. It identifies potential reversals or trend continuations.\\
5. Average True Range (ATR): Measures market volatility by calculating the average range between daily high and low prices over a specific period. \\
6. Aroon Indicators: Consist of two lines, Aroon Up and Aroon Down, which measure the time elapsed since the highest and lowest prices, respectively, within a given period. These indicators help traders identify the strength and direction of a trend. Aroon Up reaching 100 indicates a new high, while Aroon Down reaching 100 indicates a new low, suggesting a strong uptrend or downtrend. Conversely, when Aroon Up or Aroon Down falls towards 0, it suggests weakening trend momentum or potential trend reversal.\\

\subsection{Dimensional Reduction}\label{AA}
The current Quantum Hardware technology is in the developing stage, and it is highly sensitive, in turn, more prone to noise-induced errors with regard to a few of several Quantum bits {\emph{i.e., Qubits} utilized in the computing process. The Quantum advantage of handling exponentially large data with a limited number of Qubits comes with a cost of difficulty in realizing an ideal set of noise-free Qubits. Whether it is real hardware or simulation on classical hardware, one has to overcome this limitation. Hence, to overcome such limitation a minimal number of quality Qubits were involved, as a result, with the constraint of limited quality Qubits, the computing process has to be spot on. quality Qubits can not be wasted on useless parts of huge raw data, that can be ensured by doing Dimensional Reduction.

Techniques like Principal Component Analysis (PCA) and Feature selection using Quantum Annealing are employed to compress high-dimensional data and extract essential features, wherein relevant information is retained that are needed for training the prediction model. Thereby mitigating misuse of available Quantum Hardware resources so that Quantum Algorithms can operate efficiently within hardware limitations, making Quantum Computation more practical for real-world applications.

\subsubsection{Principle Component Analysis}
This is a widely used technique for dimensional reduction in data analysis. PCA transforms the original features of high-dimensional data sets into a new set of uncorrelated variables called principal components. These components are ordered based on the amount of variance they exhibit in the data. Here, the first component captures the highest variance, and the subsequent variances were captured by later ones. Since the data with high variance dictates its nature, the subset of such top principal components is selected as the PCA process. This effectively reduces the dimensionality of the data set while ensuring the most relevant information. In addition to increasing the speed of the computation, PCA offers simplification of data visualization.

\subsubsection{Quantum Annealing for Feature Selection}
The quantum version of Feature Selection is formulated as a combinatorial optimization problem known as a Quadratic Unconstrained Binary Optimization (QUBO) run on D-Wave Quantum Annealing Computers.  In this method, a feature that is selected or not is represented in the form of binary variables, which are constrained to take the value ``1"  if the feature is selected, otherwise ``0". The relevance of each feature is determined by its correlation with the target variable. As QUBO can even find solutions without enforcing the constraints, that empowers Quantum Computers over Classical counterparts. The QUBO formulation seeks to find the binary variable assignment that optimizes the objective function, which corresponds to the subset of features that best contributes to the prediction or classification task. The objective function of the QUBO is designed to maximize the relevance of selected features while minimizing redundancies and irrelevant features. The Optimization Problem of QUBO Feature Selection is described below:
\[
\text{Minimize} \quad Q(\mathbf{x}) = x^Tqx = \sum_{i=1}^{N} \sum_{j=i+1}^{N} q_{ij} x_i x_j,
\]
where $\bf x$ is vector of binary variables $x_i (\text{or} \,x_j)=\lbrace 0,1\rbrace$, and given real-valued upper triangular matrix $\mathbf{q}$ $\in$\, $\mathbf{R}^{N \times N}$ whose entries $q_{ij}$ are weights. Implementing Quantum Annealing Algorithms on D-Wave Quantum Annealer to undertake the QUBO for feature selection, the most relevant features required for the given stock market indicators can be identified \cite{vlasic2022advantage,von2021quantum}.
 
\subsection{Prediction Model}
 Comparison of stock prices on subsequent days, say on day T and day T+1, a binary classification task can be designed so that it allows predicting future stock prices. Such prediction model has to be trained first using all the extractable features. The simplest binary classification task can be defined in terms of Binary function `Change' as below:
 \[
\text{Change} = \begin{cases}
    1, & \text{if}~\, \text{Price}_\text{T} \, < \text{ Price}_\text{T+1} \\
    0, & \text{otherwise}
\end{cases}
\]
 Pertaining to training classical machine learning models, SVM, Decision Tree, Random Forest, KNN, Logistic Regression, Naive Bayes, Gradient Boosting, and XGBoost can be employed. To evaluate the impact of dimensionality reduction techniques on prediction accuracy, use of PCA and Feature Selection methods can be fruitful. PCA reduces the dimensionality of the feature space by transforming the original variables into a set of uncorrelated principal components, whereas Feature Selection aims to select the most relevant and informative features from the data set. We prepared data sets with 3, 5, and 8 features from the original data set. These new low-dimension data sets are named PCA3, PCA5, PCA8, QA3, QA5, and QA8. We utilized a diverse set of machine learning models to conduct a comprehensive comparison between the performances of PCA and Feature Selection in stock price prediction. By employing multiple models and carefully evaluating their outcomes, we aimed towards providing a thorough analysis of the effectiveness of these dimensionality reduction techniques in improving the accuracy of our predictive models. 

The Quantum Support Vector Classifier (QSVC) is an innovative extension of scikit-learn's sklearn.svm.SVC classifier, introducing the concept of quantum kernels. By integrating quantum kernels, QSVC enhances classification performance through quantum feature mapping. One such feature map is the ZZ Feature Map, which transforms classical input data into quantum states by leveraging ZZ couplings between Qubits. In QSVC, the ZZ Feature Map is applied to encode the dataset into a quantum state, exploiting quantum entanglement and correlations among qubits to capture complex relationships within the data. This enables QSVC to process and classify data in a quantum-enhanced manner, offering the potential for improved performance in challenging classification tasks. This combination of QSVC and the ZZFeature Map exemplifies the fusion of quantum computing techniques and classical machine learning, promising advancements in solving complex real-world problems.
We trained Quantum Support Vector Classifier (QSVC) models using the datasets with reduced dimensionality. The Qiskit SDK offers 4 different entanglement schemes for the ZZ-Feature Map\cite{Qiskit}. We trained the QSVM model using all four entanglement schemes namely `linear', `circular', `full' and `pairwise' and compared the accuracy achieved in each case. This allows us to explore the effect of feature maps on the accuracy of the QSVM model. 

\section{Results and Discussion}

\begin{figure*}
    \centering
    \begin{subfigure}{0.40\textwidth}
        \centering
        \includegraphics[width=\textwidth]{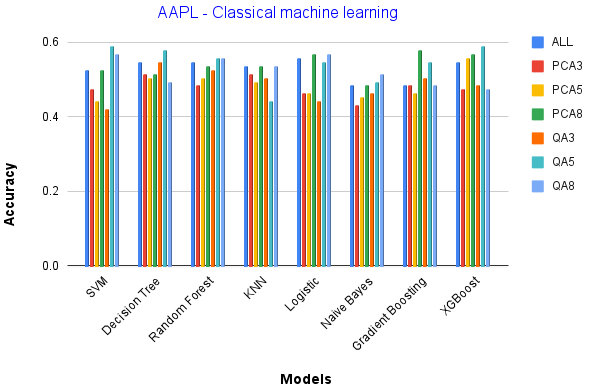}
        \caption{Apple Classical}
    \end{subfigure}%
    \begin{subfigure}{0.40\textwidth}
        \centering
        \includegraphics[width=\textwidth]{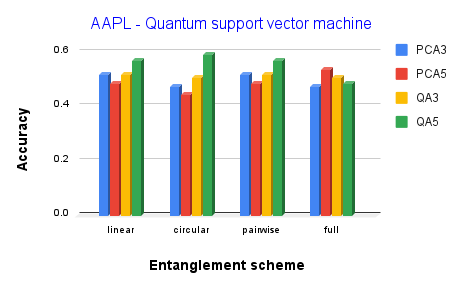}
        \caption{Apple Quantum}
    \end{subfigure}%
        
    \begin{subfigure}{0.40\textwidth}
        \centering
        \includegraphics[width=\textwidth]{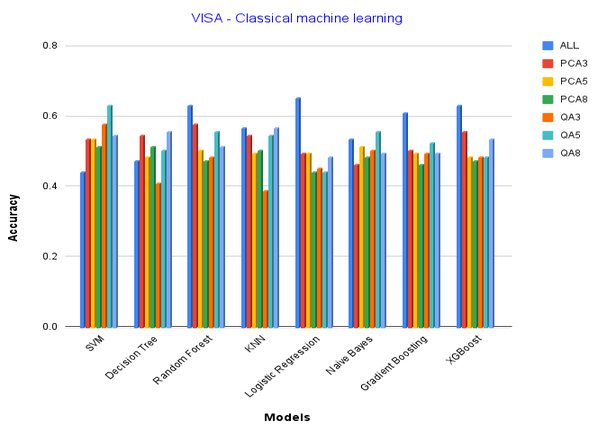}
        \caption{Visa Classical}
    \end{subfigure}
    \begin{subfigure}{0.40\textwidth}
        \centering
        \includegraphics[width=\textwidth]{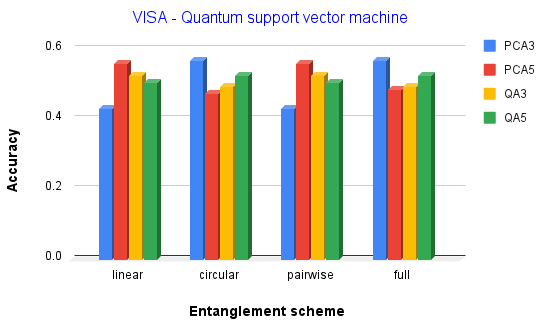}
        \caption{Visa Quantum}
    \end{subfigure}

    \begin{subfigure}{0.40\textwidth}
        \centering
        \includegraphics[width=\textwidth]{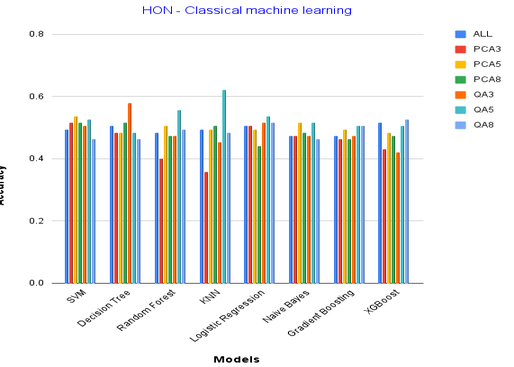}
        \caption{Honeywell Classical}
    \end{subfigure}%
        \begin{subfigure}{0.40\textwidth}
        \centering
        \includegraphics[width=\textwidth]{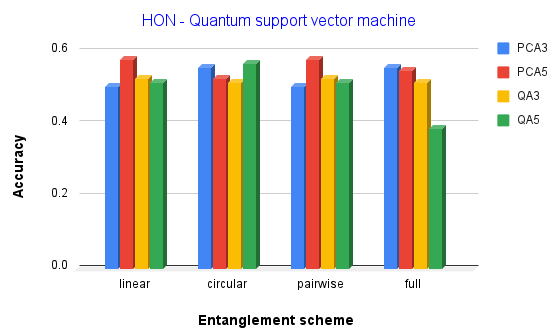}
        \caption{Honeywell Quantum}
    \end{subfigure}%

    \begin{subfigure}{0.40\textwidth}
        \centering
        \includegraphics[width=\textwidth]{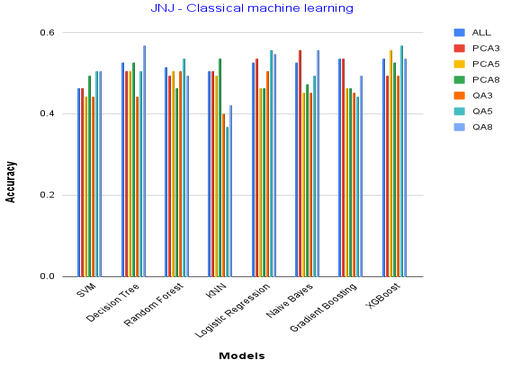}
        \caption{Johnson and Johnson Classical}
    \end{subfigure}%
    \begin{subfigure}{0.40\textwidth}
        \centering
        \includegraphics[width=\textwidth]{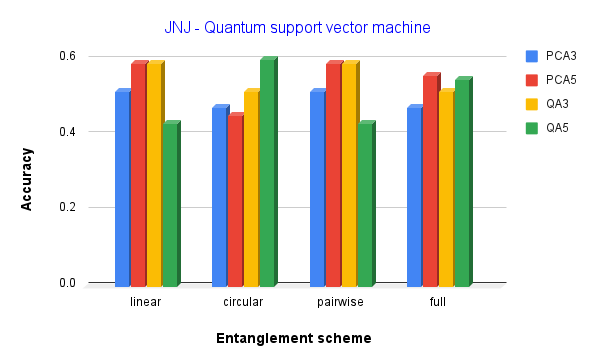}
        \caption{Johnson and Johnson Quantum}
    \end{subfigure}%
    
\label{Fig:2}
    \caption[A short caption]{Plots for Accuracy}
\end{figure*}


\begin{table*}[!ht]
    \centering
    \begin{tabular}{||c |c |c|c||} 
        \hline
        \textbf{Model}  & \textbf{Dimensionality Reduction} & \textbf{AVG Accuracy} & \textbf{AVG F-Score} \\ [0.5ex] 
        \hline\hline
        Classical Machine Learning & None & 56.58\% & 58.76\%  \\
        \hline
        Classical Machine Learning & PCA-3 & 54.21\% & 56.81\%  \\
        \hline
        Classical Machine Learning & PCA-5 & 55\% & 58.83\%  \\
        \hline
        Classical Machine Learning & PCA-8 & 53.68\% & 57.71\% \\
        \hline
        Classical Machine Learning & Quantum Annealing-3 & 55.26\% & 52.81\% \\
        \hline
        Classical Machine Learning & Quantum Annealing-5 & \textbf{60.26\%} & \textbf{62.24\%}  \\
        \hline
        Classical Machine Learning & Quantum Annealing-8 & 55.79\% & 57.22\%  \\
        \hline

        QSVM  & PCA-3 & 53.94\% & 58.08\%  \\
        \hline
        QSVM  & PCA-5 & 56.58\% & 59.79\%  \\
        \hline
        QSVM  & Quantum Annealing-3 & 53.94\% & 54.04\%  \\
        \hline
        QSVM  & Quantum Annealing-5 & 55.78\% & 57.45\%  \\
        \hline

    \end{tabular}
    \caption{Average Accuracy }
    \label{table:avg-acc}
\end{table*}

\begin{table*}[!ht]
    \centering
    \begin{tabular}{||c |c |c |c|c||} 
        \hline
        \textbf{Model} &\textbf{Entanglement Scheme} & \textbf{Dimensionality Reduction} & \textbf{Accuracy} & \textbf{F-Score} \\ [0.5ex] 
        \hline\hline
        XG Boost &None & None & 51.58\% & 58.18\% \\ 
        \hline
        SVM & None & PCA-3 & 51.58\% & 58.18\% \\
        \hline
        SVM & None & PCA-5 & 53.68\% & 60.71\% \\
        \hline
        SVM & None & PCA-8 & 51.58\% & 57.40\% \\
        \hline
        
        Decision Tree & None & Quantum Annealing-3  & 57.89\% & 62.26\% \\
        \hline        
        K- Nearest Neighbours & None & Quantum Annealing-5  & \textbf{62.10\%} & \textbf{68.96\%} \\
        \hline 	
        XGBoost	& None & Quantum Annealing-8  &52.61\% & 57.94\% \\
        \hline        
        Quantum SVM & Linear & Quantum Annealing-3  & 52.63\% & 55.45\% \\
        \hline
        Quantum SVM & Full & PCA-3  & 55.78\% & 61.81\% \\
        \hline
        Quantum SVM & Circular & Quantum Annealing-5  & 56.84\% & 61.68\% \\
        \hline      
        Quantum SVM & Circular & Quantum Annealing-5  & 56.84\% & 61.68\% \\
        \hline
        Quantum SVM & Linear & PCA-5  & 57.89\% & 65.51\% \\
        \hline
        Quantum SVM & Full & PCA-5  & 57.89\% & 65.51\% \\  [1ex] 
        \hline     
        
        \hline
    \end{tabular}
    \caption{Best Models for Honeywell dataset }
    \label{table:2}
\end{table*}

\begin{table*}[!ht]
    \centering
    \begin{tabular}{||c |c |c |c|c||} 
        \hline
        \textbf{Model} &\textbf{Entanglement Scheme} & \textbf{Dimensionality Reduction} & \textbf{Accuracy} & \textbf{F-Score} \\ [0.5ex] 
        \hline\hline
        Gradient Boosting &None & None & 53.68\% & 56\% \\ 
        \hline
        Naive Bayes & None & PCA-3 & 55.79\% & 59.61\% \\
        \hline
        Decision Tree & None & PCA-5 & 56.84\% & 58.58\% \\
        \hline
        KNN & None & PCA-8 & 53.68\% & 56.86\% \\
        \hline
        Random Forest & None & Quantum Annealing-3  & 50.52\% & 50.52\% \\
        \hline        
        XG Boost & None & Quantum Annealing-5  & 56.84\% & 57.73\% \\
        \hline
        Decision Tree & None & Quantum Annealing-8  & 56.84\% & 53.93\% \\
        \hline

        Quantum SVM & Linear & Quantum Annealing-3  & 58.94\% & 63.55\% \\
        \hline
        Quantum SVM & Pairwise & Quantum Annealing-3  & 58.94\% & 63.55\% \\
        \hline
        Quantum SVM & Linear & PCA-3  & 51.58\% & 55.77\% \\
        \hline
        Quantum SVM & Pairwise & PCA-3  & 51.58\% & 55.77\% \\
        \hline
        
        Quantum SVM & Linear & PCA-5  & 58.94\% & 58.94\% \\
        \hline  
        Quantum SVM & Pairwise & PCA-5  & 58.94\% & 58.94\% \\
        \hline      
        Quantum SVM & Circular & Quantum Annealing-5  & \textbf{60\%} & \textbf{56.82\%} \\[1ex] 
        \hline     
        
        \hline
    \end{tabular}
    \caption{Best Models for Johnson \& Johnson dataset }
    \label{table:3}
\end{table*}

\begin{table*}[!ht]
    \centering
    \begin{tabular}{||c |c |c |c|c||} 
        \hline
        \textbf{Model} &\textbf{Entanglement Scheme} & \textbf{Dimensionality Reduction} & \textbf{Accuracy} & \textbf{F-Score} \\ [0.5ex] 
        \hline\hline
        Logistic Regression &None & None & 55.79\% & 57.15\% \\ 
        \hline
        KNN & None & PCA-3 & 51.58\% & 54.90\% \\
        \hline
        XG Boost & None & PCA-5 & 55.78\% & 55.31\% \\
        \hline
        SVM & None & PCA-8 & 57.89\% & 59.18\% \\
        \hline
        Decision Tree & None & Quantum Annealing-3  & 54.73\% & 49.41\% \\
        \hline        
        SVM & None & Quantum Annealing-5  & \textbf{58.94\%} & \textbf{61.38\%} \\
        \hline
        SVM & None & Quantum Annealing-8  & 56.84\% & 60.19\% \\
        \hline
        Logistic Regression & None & Quantum Annealing-8  & 56.84\% & 60.19\% \\
        \hline
        
        Quantum SVM & Linear & Quantum Annealing-3  & 51.58\% & 48.89\% \\
        \hline
        Quantum SVM & Pairwise & Quantum Annealing-3  & 51.58\% & 48.89\% \\
        \hline
        Quantum SVM & Linear & PCA-3  & 51.58\% & 53.06\% \\
        \hline
        Quantum SVM & Pairwise & PCA-3  & 51.58\% & 53.06\% \\
        \hline
        
        Quantum SVM & Full & PCA-5  & 53.68\% & 55.10\% \\
        \hline      
        Quantum SVM & Circular & Quantum Annealing-5  & 58.94\% & 58.06\% \\[1ex] 
        \hline     
        
        \hline
    \end{tabular}
    \caption{Best Models for Apple dataset }
    \label{table:1}
\end{table*}

\begin{table*}[!ht]

    \centering
    \begin{tabular}{||c |c |c |c|c||} 
        \hline
        \textbf{Model} &\textbf{Entanglement Scheme} & \textbf{Dimensionality Reduction} & \textbf{Accuracy} & \textbf{F-Score} \\ [0.5ex] 
        \hline\hline
        Logistic Regression &None & None & 65.26\% & 63.73\% \\ 
        \hline
        Random Forest & None & PCA-3 & 57.89\% & 54.54\% \\
        \hline
        SVM & None & PCA-5 & 53.68\% & 60.71\% \\
        \hline
        SVM & None & PCA-8 & 51.57\% & 57.40\% \\
        \hline
        SVM & None & Quantum Annealing-3  & 54.89\% & 52.31\% \\
        \hline        
        SVM & None & Quantum Annealing-5  & \textbf{63.15\%} & \textbf{61.53\%} \\
        \hline
        KNN & None & Quantum Annealing-8  & 56.84\% & 56.84\% \\
        \hline
        Quantum SVM & Linear & Quantum Annealing-3  & 52.63\% & 48.28\% \\
        \hline
        Quantum SVM & Linear & Quantum Annealing-3  & 52.63\% & 48.28\% \\
        \hline
        Quantum SVM & Circular & PCA-3  & 56.84\% & 61.68\% \\
        \hline
        Quantum SVM & Full & PCA-3  & 56.84\% & 61.68\% \\
        \hline
        Quantum SVM & Linear & PCA-5  & 55.78\% & 59.61\% \\
        \hline      
        Quantum SVM & Pairwise & PCA-5  & 55.78\% & 59.61\% \\
        \hline      
        Quantum SVM & Circular & Quantum Annealing-5  & 52.63\% & 52.63\% \\
        \hline
        Quantum SVM & Full & Quantum Annealing-5  & 52.63\% & 52.63\% \\[1ex] 
        \hline     
    
    \end{tabular}
    \caption{Best Models for Visa dataset }
    \label{table:4}
\end{table*}

In this study, we looked into the applications of Quantum Computing in the financial domain, with a focus on stock price prediction and binary classification of decreasing or increasing stock prices for four different companies, namely, Apple, Visa, Johnson \& Jonson and Honeywell. We compared various Quantum Computing techniques with their classical counterparts in terms of accuracy (\emph{or}, average accuracy) and F-score of the prediction model (see Fig.2). The F-score is a metric used to evaluate the performance of a given Machine Learning Model, wherein it is evaluated by combining precision and recall.

The results from our experiments provided valuable insights into the performance of Quantum Computing methods for financial analysis (refer Tables I-V). Feature selection using quantum annealing demonstrated its ability to extract the most relevant features from financial data more effectively than PCA, presenting Quantum Annealing as a promising approach for feature selection tasks in finance. This finding is significant, as the identification of relevant features from a wide variety of stock market indicators plays a critical role in the accuracy of prediction models as well as in the decision-making processes.
However, in the binary classification task, we found that the quantum support vector machine (QSVM) did not exhibit a significant advantage over the classical support vector machine (SVM) for the given data sets. This outcome suggests that the quantum advantage might not be evident in all financial analysis scenarios, and further investigation is required to explore other potential applications of QSVM in finance. We implemented the QSVC model for data sets with 8 features, however, these models failed to train due to the insufficient computational power, which is required for simulating the quantum circuits of the feature maps. This further emphasises the importance of using dimensionality reduction for Quantum Machine Learning pipelines (refer Fig.1).

The data-set's diversity from four distinct domain companies allowed us to explore the applicability of quantum techniques across various sectors, and identify potential domain-specific trends. This holistic analysis provides valuable information for investors and financial analysts seeking to leverage Quantum Computing to enhance their decision-making.


\section{Conclusion}

In conclusion, our investigation on the application of Quantum Computing in the financial domain has provided valuable insights with regards to its potential and limitations. Quantum Annealing exhibited promise in the feature selection task, surpassing PCA in extracting relevant patterns from financial data. However, the quantum advantage was not evident in binary classification, as QSVM did not outperform classical SVM by a significant margin. Nevertheless, the diverse dataset enabled us to explore quantum techniques' applicability across various sectors, identifying domain-specific trends. While Quantum Computing shows promise, its full potential in financial analysis requires further research and development. Overall, this study contributes to the understanding of quantum computing's role in the finance domain and provides a foundation for future investigations in quantum-enhanced financial analysis and decision-making.

\bibliographystyle{unsrt} 

\end{document}